\documentclass[%
 preprint,
 amsmath,amssymb,
 aps,
]{revtex4-1}

\usepackage{graphicx}% Include figure files
\usepackage{dcolumn}% Align table columns on decimal point
\usepackage{bm}% bold math
\usepackage{upgreek}
 
\begin{document}

%\preprint{APS/123-QED}

\title{Ultra-high-$Q$ phononic resonators on-chip at cryogenic temperatures}% Force line breaks with \\
%\thanks{A footnote to the article title}%

\author{Prashanta Kharel}
%  \altaffiliation[Also at ]{Physics Department, XYZ University.}%Lines break automatically or can be forced with \\
\email{prashanta.kharel@yale.edu}
\author{Yiwen Chu}%
\author{Michael Power}
\author{William H. Renninger}
\author{Robert J. Schoelkopf}
\author{Peter T. Rakich}
\email{peter.rakich@yale.edu}
 % \email{Second.Author@institution.edu}
\affiliation{%
 Department of Applied Physics, Yale University, New Haven, CT 06511\\
%  This line break forced with \textbackslash\textbackslash
}

% \collaboration{MUSO Collaboration}%\noaffiliation

% \author{Charlie Author}
%  \homepage{http://www.Second.institution.edu/~Charlie.Author}
% \affiliation{
%  Second institution and/or address\\
%  This line break forced% with \\
% }%
% \affiliation{
%  Third institution, the second for Charlie Author
% }%
% \author{Delta Author}
% \affiliation{%
%  Authors' institution and/or address\\
%  This line break forced with \textbackslash\textbackslash
% }%

% \collaboration{CLEO Collaboration}%\noaffiliation

\date{\today}% It is always \today, today,
             %  but any date may be explicitly specified

\begin{abstract}

Long-lived, high frequency phonons are valuable for applications ranging from optomechanics to emerging quantum systems. For scientific as well as technological impact, we seek high-performance oscillators that offer a path towards chip-scale integration. Confocal bulk acoustic wave resonators have demonstrated an immense potential to support long-lived phonon modes in crystalline media at cryogenic temperatures. So far, these devices have been macroscopic with cm-scale dimensions. However, as we push these oscillators to high frequencies, we have an opportunity to radically reduce the footprint as a basis for classical and emerging quantum technologies. In this paper, we present novel design principles and simple fabrication techniques to create high performance chip-scale confocal bulk acoustic wave resonators in a wide array of crystalline materials. We tailor the acoustic modes of such resonators to efficiently couple to light, permitting us to perform a non-invasive laser-based phonon spectroscopy. Using this technique, we demonstrate an acoustic $Q$-factor of 28 million (6.5 million) for chip-scale resonators operating at 12.7 GHz (37.8 GHz) in crystalline $z$-cut quartz ($x$-cut silicon) at cryogenic temperatures.

% However, as we push these to radically reduced the size 
% These strategies could be enabling 

%while phononic crystal based device strategies offer one solution, the desin and realization  of such systems is often non-trivial

%Long lived phonon modes have applications in fields ranging from optomechanics, electro-mechanics and emerging hybrid quantum phononic systems. Plano-convex bulk crystalline phonon resonators have been demonstrated to support such long-lived phonons at cryogenic temperatures where optomechanical and quantum systems are routinely operated. In an effort to marry such high-Q phononic resonator technologies with existing optomechanical and quantum systems, we use simple, materials-agnostic microfabrication technique to create ultra-high Q-factor phononic resonators on-chip. We engineer tightly confined phonon modes that have good acousto-optic overlap with laser fields, permitting non-invasive laser-based phonon mode spectroscopy. Using this technique, we demonstrate ultra-high Q-factor - 28 million and .6 million - phononic resonators on-chip at high-frequencies - 12.7 GHz and 37.8 GHz- in crystalline z-cut quartz and x-cut silicon respectively. These strategies and techniques for obtaining high f$\cdot$Q-products- $3.6\times10^{17}$ and $2.5 \times 10^{17}$- on-chip for quartz and silicon respectively could enable classical and quantum devices with greatly enhanced performance. 
\end{abstract}

\pacs{Valid PACS appear here}% PACS, the Physics and Astronomy
                             % Classification Scheme.
%\keywords{Suggested keywords}%Use showkeys class option if keyword
                              %display desired
\maketitle

\section{INTRODUCTION}
%\tableofcontents
Acoustic-wave technologies have become indispensable for everything from classical signal processing \cite{lin1998microelectromechanical} to precision metrology \cite{miklos2001application}.  Rapid advancements in quantum optics, optomechanics, and circuit quantum electrodynamics have recently spurred interest in phonons as the basis for  emerging quantum technologies \cite{blencowe2004quantum, aspelmeyer2014cavity,o2010quantum, gustafsson2014propagating, Chu199}. In these systems, phonons become coherent carriers of information and can also be utilized to mediate interactions between different types of excitations (such as optical photons, microwave, and defect centers) \cite{poot2012mechanical, palomaki2013coherent, o2010quantum, Chu199, andrews2015quantum, soykal2011sound}. In this context, we seek high quality-factor ($Q$) phonon modes at high frequencies ($f$), making the $f\! \cdot \! Q$-product a key figure of merit \cite{aspelmeyer2014cavity, poot2012mechanical}. Such high frequency (GHz) phonons are easily cooled to their quantum ground states and can be used to store quantum states for extended periods of time. Low temperature operation has the added benefit that phonon lifetimes are radically enhanced in pristine crystalline media. As a basis for emerging quantum technologies \cite{Chu199}, we seek high performance phononic resonators that offer a path towards chip-scale integration.

While there are many promising approaches to size reduction for mechanical resonators \cite{aigner2003mems}, the task of achieving high $f\! \cdot \! Q$ products within a small package at cryogenic temperatures introduces a unique set of challenges. To dramatically extend phonon lifetimes, we seek to eliminate extrinsic sources of loss. In this regard, phononic crystal-based device strategies offer an intriguing solution \cite{olsson2008microfabricated,eichenfield2009optomechanical,safavi2014two,hong2017hanbury}, as they can theoretically eliminate external loss channels through the formation of complete phononic bandgaps. Bulk acoustic wave (BAW) resonators offer a complementary path to high performance at cryogenic temperatures. For instance, record $f\! \cdot \! Q$ products ($1.6 \times 10^{18}$) have been demonstrated at microwave (200 MHz) frequencies using confocal BAW resonators that mitigate extrinsic losses by trapping the phonon modes within the bulk of a pristine crystal \cite{galliou2013extremely}. However, because these devices are typically designed for operation at relatively low frequencies (5-100 MHz), they have historically been relatively large (centimeter-scale) \cite{besson1977new}. Fortunately, as we scale to higher phonon frequencies, smaller acoustic wavelength permits radical reductions in size.  These resonators are comparatively simple to fabricate and have great versatility; individual resonators support high-$Q$ phonon modes over a wide range of frequencies (1-100 GHz), and can be formed from an array of different materials.

\begin{figure} 
\centering
\includegraphics[width=0.6\textwidth]{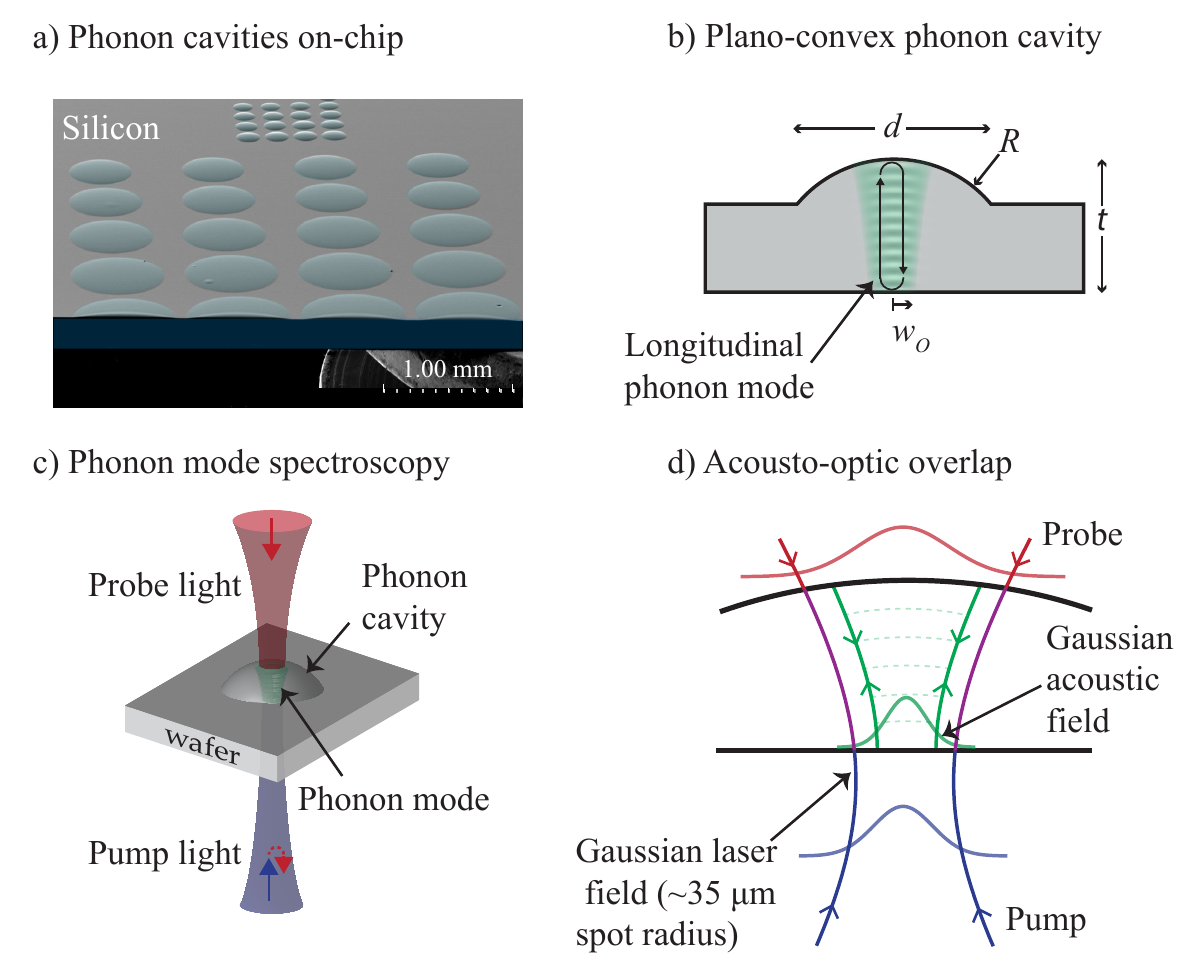}
\caption{\label{fig1} a) False-color scanning electron microscope image of arrays of plano-convex bulk acoustic wave resonators microfabricated on a silicon chip. Phonon cavity geometry can be easily tailored for resonators fabricated on-chip. b) These resonators have a plano-convex geometry that permits tight confinement of standing-wave longitudinal phonon modes near the center of the device (acoustic waist radius $w_o < 40 \ \upmu$m). c) These longitudinal acoustic modes couple to counter-propagating optical fields through photoelastic coupling, which is used to perform non-invasive laser-based spectroscopy of phonon modes at GHz frequencies. d) For sensitive optomechanical spectroscopy, we seek to maximize the acousto-optic overlap with Gaussian laser fields with spot sizes of $\sim 35 \ \upmu$m.}
\end{figure}

In this paper, we present novel design principles and simple fabrication techniques to create high performance BAW resonators on-chip. Through microfabrication of such resonators in two different materials, we demonstrate that these techniques can be adapted to create BAW resonators in a wide array of materials. We engineer the modes of these resonators to produce efficient light-sound coupling using the instrinsic photoelastic response, which is present in all transparent media. This permits us to perform non-invasive laser based phonon spectroscopy at high frequencies (10-40 GHz). Using these device strategies and measurement approach, we demonstrate $Q$-factors of 28 million (6.5 million) for 12.7 GHz (37.8 GHz) phonon modes of  microfabricated resonators in $z$-cut quartz ($x$-cut silicon). Remarkably, these chip-scale BAW resonators, with $>$1000-fold reduction in device volumes, exhibit $f\! \cdot \! Q$ products on par with previously demonstrated values in macroscopic (cm-scale) BAW resonators \cite{renninger2017bulk}.

\section{Device Design}

Our device consists of a microfabricated plano-convex BAW resonator on-chip (see Fig. 1a), which we call confocal High-Overtone Bulk Acoustic Resonator (cHBAR). Microfabrication permits creation of compact devices with diameters ranging from tens of microns up to a few millimeters. Longitudinal acoustic phonons are trapped within this system as the acoustic wave reflects from both top and bottom surfaces of the substrate, forming standing-wave phonon modes (see Fig. 1b). This plano-convex design mitigates acoustic diffraction and produces confinement of acoustic energy in the transverse dimension (typical acoustic beam radius is $<$40 $\upmu$m). Since these acoustic modes live primarily in the bulk medium, they have greatly reduced surface interactions. 

The response of the phonon field is detected using a non-invasive laser-based spectroscopy; stimulated energy transfer between counter-propagating light fields occurs as the detuning between the two light fields is swept near the Brillouin frequency  (see Fig. 1c) \cite{renninger2017bulk}. This energy transfer spectrum gives us information about the frequencies and lifetimes of the phonons. To permit efficient optical access to the resonator modes, we also tailor the plano-convex geometry to enhance optomechanical coupling with the incident Gaussian laser fields of spot sizes of $\sim 35\ \upmu$m (see Fig. 1d). Before we explore this optomechanical coupling in greater detail, we explain the design principles and fabrication methods used to create chip-scale cHBAR.

Since the wavelengths ($\lambda_\text{ph} < 1 \  \upmu$m) of these high frequency phonons are much smaller than the system dimensions (hundreds of microns), the acoustic wave propagation becomes reminiscent of optical beam propagation in the paraxial limit. We show that these plano-convex phonon resonators support high $Q$-factor mode families, with Hermite-Gaussian-like mode profiles (See Supplementary Note for details); the acoustic mode profile is engineered to have good acousto-optic overlap with the Gaussian laser fields. Next, we adopt established methods from optics to design stable, high-$Q$ cHBAR.

Within the framework of Gaussian optical resonator design, we expect only certain radii of curvatures to form stable cavities; in a stable optical cavity, transverse spatial confinement occurs because the reflections from the resonator surfaces compensate for effects of diffraction.  Stability criteria for a Fabry-P{\'e}rot optical cavity in vacuum consisting of two mirrors with radii of curvatures $R_1$ and $R_2$ separated by a distance of $L$ is given by $0 \leq g_1 g_2 \leq 1 ,$ where the stability parameter is defined as $g_i = 1-L/R_i$ ($i=1, 2)$ \cite{siegman1986lasers}. In the context of acoustics, correctly formulated stability parameters $g_1$ and $g_2$ must account for the anisotropy of elastic constants. In contrast to slowness surfaces for optical waves propagating in vacuum, acoustic slowness surfaces are not necessarily symmetric or even parabolic \cite{royer2000elastic}; as a result, acoustic beam propagation in crystalline media can be non-trivial. However, we can greatly simplify the acoustic resonator design problem by choosing crystalline axes about which the dispersion surfaces are parabolic and symmetric to first order. In this case, we can formulate stability criteria that closely mirror laser beam optics (See Supplementary Note for details). The stability parameters for plano-convex phononic cavities formed using $z$-cut quartz and $x$-cut silicon are simply given by

%Within the framework of Gaussian beam resonator design, we expect only certain ranges of $R$ to produce stable (high $Q$-factor) phononic resonators for a given thickness $t$ \cite{siegman1986lasers}. In an unstable phononic resonator, the intracavity acoustic beam grows without limit giving rise to leaky modes. However, in a stable resonator, transverse spatial confinement of the phonon beam occurs as the acoustic beam is periodically refocused. Stability criterion analogous to that of optical cavity i.e. $0 \leq g_1 g_2 \leq 1 $ can be derived for acoustic cavities along certain crystalline axes. For optical cavities, the standard stability parameter is given by $g_i=1-t/R_i$, with $i = 1,2$. In the context of acoustics, correctly formulated stability parameters $g_1$ and $g_2$ must account for the anisotropic nature of acoustic wave propagation. We greatly simplify this acoustic design problem  by choosing crystalline axes such that dispersion surfaces are parabolic to first order. In this case, we can formulate stability criteria that closely mirror laser beam optics (See Appendix).  Therefore, the stability parameters for plano-convex phononic cavities formed along $z$-cut quartz and $x$-cut silicon are simply given by
\begin{equation}
    g_1 = 1, g_2 = 1- \frac{t}{\chi R},
\end{equation}
where $t$ is the thickness of the wafer, $R$ is the radius of curvature of the convex surface, and $\chi$ is an ``anisotropy-constant" that includes the effect of propagation of acoustic beam in an anisotropic medium. For acoustic beam propagation perpendicular to the $z$-cut face of quartz and $x$-cut face of silicon, $\chi$ can be calculated analytically yielding $\chi_{\text{Si}}= 0.6545$ and $\chi_{\text{Quartz}}=0.5202$ (See Supplementary Note for details). Therefore, the range of radius of curvatures that can produce stable plano-convex phonon cavities is $0 \leq 1- t/ (\chi R) \leq 1 $ or equivalently $R\geq  t/\chi$.  For the design of phonon  cavities  along  crystalline  axes  that produce non-trivial  dispersion  surfaces, more sophisticated methods such as the numerical  acoustic beam propagation techniques discussed in Ref.  \cite{renninger2017bulk} must be used.
%The results of the numerical acoustic beam propagation (stability criteria, phonon mode waist, and the frequency spacing for the higher order modes) matches well with the analytically derived values for the phonon cavities on both x-cut silicon and z-cut quartz.

In addition to forming a stable phonon cavity, we choose the radius of curvature $R$ to enhance the acousto-optic coupling. Since the acousto-optic coupling depends on the overlap integral between the optical and acoustic modes \cite{renninger2017bulk}, we seek to maximize coupling for the fundamental acoustic mode by matching the acoustic beam waist to the optical beam waist. The acoustic waist radius, $w_o$, at the planar surface can be expressed in terms of $R$ as  
\begin{equation} \label{waistRadius}
    w_o^2 = \frac{t  \lambda_{\text{ph}}}{\chi\pi}  \sqrt{\frac{g_1g_2 (1-g_1g_2)}{(g_1+g_2-2g_1g_2)^2}},
\end{equation}
where $\lambda_{\text{ph}}$ is the wavelength of the phonon mode.  For instance, a 1 mm thick plano-convex phonon cavity in z-cut quartz with $R = 65$ mm supports a 12.66 GHz acoustic mode having $w_o= 39.6 \ \upmu$m. Therefore, by changing $R$, we can tailor the acoustic mode so that it couples efficiently to a focused laser beam of radius $\sim$35 $\upmu$m used in our experiments. Finally, we chose the diameter of the phonon cavity, $d$, to be much larger than the phonon beam waist, $w_o$ (See Fig. \ref{fig1} b)) . This ensures that the exponential tails of the acoustic Gaussian beam are vanishingly small where the convex surface terminates, meaning the diffractive (or anchoring losses) are negligible. For instance, assuming all the energy outside the convex surface is lost to diffractive losses, for $d/w_o = 5$, we still find that the $Q$-factor limit due to this loss mechanism would be 7 billion (15 billion) for the phonon cavities fabricated on quartz (silicon) (See Supplementary Note for details). For a phonon cavity with given thickness and radius of curvature, it is easy to see from Eq. (\ref{waistRadius}) that phonons at higher frequencies have smaller waist radius (i.e. $w_o \propto 1/\sqrt{f})$. Therefore, as we seek operation at high frequencies, the phonon mode volume shrinks, permitting us to fabricate smaller phononic devices. 

\section{fabrication}
\begin{figure} 
\centering
\includegraphics[width=0.95\textwidth]{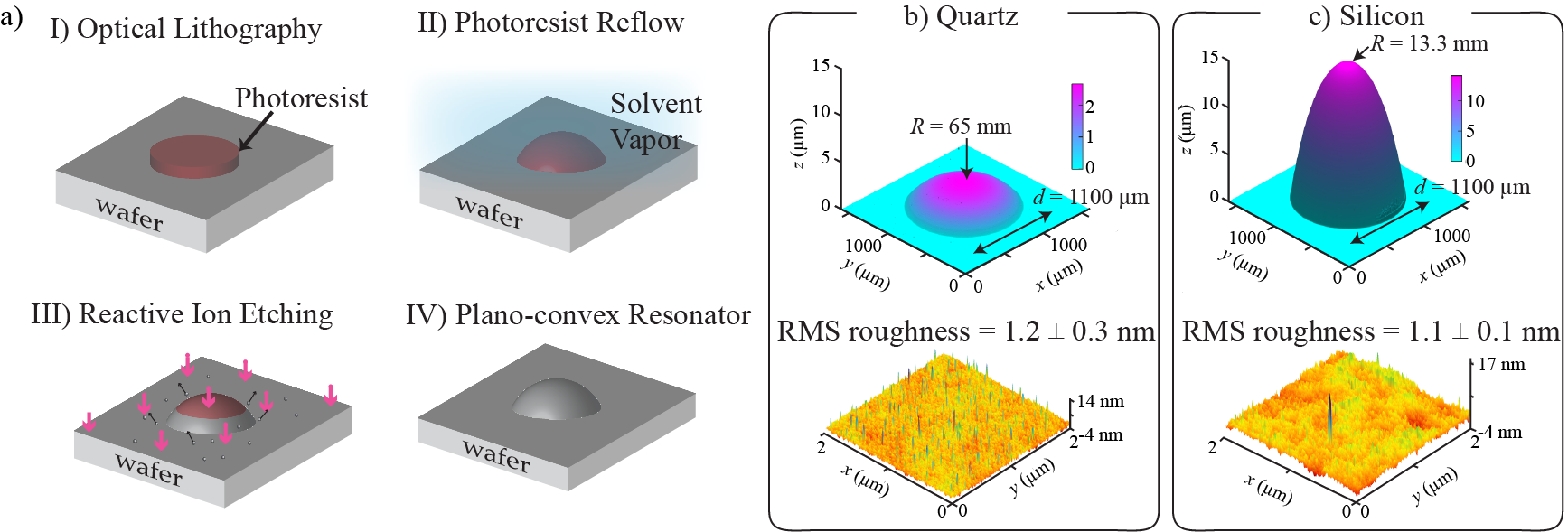}
\caption{\label{fig2} a) Microfabrication steps taken to fabricate plano-convex phononic resonators on-chip. I) Standard optical lithography is used to write cylindrical photoresist patterns. II) This photoresist cylinder is reflowed into a hemisphere using solvent-vapor reflow technique. III) Reactive ion etching (RIE) is used to transfer the hemispherical pattern onto the substrate. IV) After completely etching away the photoresist, we get a plano-convex cavity that supports long-lived phonon modes. Measured 3D-surface profiles of the phonon cavities in b) $z$-cut quartz and c) $x$-cut silicon after RIE show excellent surface quality. The difference in height (and hence the radius of curvature) is a result of difference in etch selectivity of the substrate over photoresist (ratio of etch rate of substrate to etch rate of photoresist). Using atomic force microscopy (AFM), we measure $\sim$ 1 nm root mean square surface roughness for phononic cavities on both quartz and silicon. }
\end{figure}

Next, we give an overview of the fabrication steps used to fabricate these phonon cavities. We developed simple microfabrication techniques to create high preformance phononic oscillators on-chip by leveraging strategies used to fabricate optical micro-lenses \cite{popovic1988technique,eisner1996transferring, li2004fabrication}. However, to achieve low loss phonon modes at high frequencies (having $\lambda_\text{ph} \lessapprox 500 $ nm), we optimized the fabrication process to yield resonators having excellent surface finish. Standard optical lithography allowed us to print cylindrical photoresist patterns which are transformed into hemispheres using a solvent vapor reflow technique (See Fig.  $\ref{fig2}$a). Photoresist structures became less viscous after absorbing solvent vapor, allowing surface tension to form photoresist hemispheres having excellent surface roughness ($\sim$ 1 nm).  This reflow also permits us to form photoresist hemispheres having large radii of curvature (tens of mm).  Additionally, the center height of the photoresist hemispheres after solvent vapor reflow is not dependent on the substrate material and is relatively insensitive to the photoresist diameters \cite{emadi2009vertically}. In this way, we can change the radius of curvature by independently changing the thickness or the diameter of the photoresist cylinders. 

After the reflow process, hemispherical photoresist patterns are imprinted on the substrate material by completely etching away the photoresist using reactive ion etching (RIE). The ratio of etching rate of photoresist to the etching rate of substrate material (also known as etch selectivity), determined the final radius of curvature of the phonon cavities. Reactive ion etch parameters were optimized to ensure excellent surface roughness after the etch. 

In what follows, we outline detailed fabrication steps used to make plano-convex phonon cavities in crystalline quartz and silicon on-chip. The fabrication process begins by creating photoresist hemispheres. We start with a double-sided polished 1 mm thick $z$-cut quartz wafer ($>$ A grade, 99.997$\%$ pure). To eliminate the possibility of organic contaminants and adsorbates, the wafer is oxygen plasma cleaned for 3 minutes at RF-power of 300 Watts and a pressure of 300 mTorr. We then spin coated a 5.5 $\upmu$m thick layer of photoresist (AZP 4620) on the wafer. A post-bake at 110 $^{\circ}$C for 2 minutes was performed to harden the photoresist. A lithographic photomask was used to define circular structures during UV exposure (400 mJ/cm$^2$ at 405 nm wavelength). The exposed photoresist is developed using 1:4 AZ400K:water developer solution. The photoresist cylinders are then vapor primed with resist adhesion promoter hexamethyldisilizane (HMDS) for 15 minutes. Solvent vapor reflow of these photoresist cylinders is accomplished using polypropylene glycol monomethyl ether acetate (PGMEA) solvent. The solvent is heated in a closed chamber at 55 $^{\circ}$C with the wafer placed upside down (not touching the liquid) at 60 $^{\circ}$C until the photoresist has completely reflowed into hemispheres. After the reflow, the wafer is first baked at 90 $^{\circ}$C for 1 minute to harden the photoresist and the temperature is gradually increased to 125 $^{\circ}$C over the course of 15 minutes to get rid of excess solvent. The center height of the photoresist hemispheres after the reflow process is approximately 10.5 $\upmu$m. For this resist thickness, by simply changing the diameter of the photoresist hemispheres from 50 $\upmu$m to $1.5$ mm, we can vary the radii of curvature of these photoresist hemispheres from approximately hundreds of microns to 30 mm.
 
A slow reactive ion etch using SF$_6$ and Ar gases with 4 sccm and 14 sccm flow rates respectively at a low chamber pressure of 4 mTorr and a bias voltage of 370 V is used to etch away the photoresist completely. In the process, a slow erosion ($\sim$ 35 nm/min) of the substrate occurs as shown in Fig. \ref{fig2} a-III. This results in phonon cavities with excellent surface quality. The etch selectivity of photoresist over quartz of 3.8 results in hemispheres in quartz with center height of approximately 2.7 $\upmu$m. Because of the etch-selectivity, we can fabricate phonon cavities with radius of curvatures as large as 110 mm in quartz by simply changing the diameter of the photoresist. However, even larger radii of curvature can be made by using thinner photoresist. Finally, the wafer  is cleaned in piranha solution (3:1 sulphric acid:hydrogen peroxide) for 2 minutes to get rid of any organic contaminants before optical measurements. 

The fabrication steps of phonon cavities on silicon are similar to those outlined for quartz above; only the reactive ion etching parameters differ. We create photoresist hemispheres on a 500 $\upmu$m thick double-side-polished float-zone grown $x$-cut silicon wafer (resistivity $>$ 1000 $\Omega \cdot$cm) using the same optical lithography and solvent vapor reflow outlined for quartz. The photoresist hemispheres are then reactive ion etched using SF$_6$ and O$_2$ gases with 5 sccm and 2 sccm flow rates, respectively, at a chamber pressure of 10 mTorr and a bias voltage of 394 V. A slow erosion ($\sim$100 nm/min) of the substrate gives excellent surface quality. The etch selectivity of 0.74 results in hemispheres in silicon with center height of approximately 14.2 $\upmu$m. Surface passivation of silicon using piranha etch and diluted hydrofluoric acid dip as outlined in Ref. \cite{borselli2006measuring} is performed before optical measurements. 

The results of these fabrication processes are first characterized using a 3D surface profilometer (Zygo Nexview) as seen in Fig. \ref{fig2} b-c. A hemispherical surface fits well to these plano-convex phonon cavities with diameters of 1100 $\upmu$m. This fitting allows us to determine the radius of curvature of 65 $\pm$ 1 mm (13.3 $\pm$ 0.3 mm) for resonators in quartz (silicon). This difference in the radius of curvatures for the same diameter cavities in two different substrates is a result of etch selectivity differences during RIE process. We chose 65 mm radius of curvature phonon cavities in quartz for our optical measurements because, as discussed before, this results in enhanced acousto-optic overlap  with the Gaussian laser fields. We chose 13.3 mm radius of curvature phonon cavities in silicon because, given low etch selectivity of 0.74, this was the largest radius of curvature we could obtain without significant deviation from a hemispherical surface. We observed that the photoresist hemispheres with large diameters ($>1.1$ mm) tend to have asymmetry in the convex surfaces. The surface roughness is measured using an atomic-force microscope (Bruker Dimension Fastscan AFM). We measure a root mean square roughness of 1.2 $\pm$ 0.3 nm (1.1 $\pm$ 0.1 nm) for the etched surface in quartz (silicon). These fabrication results show promising features that hint at phonon cavities that should support long-lived phonons with minimal scattering losses.
\begin{figure}
\includegraphics[width=0.90\textwidth]{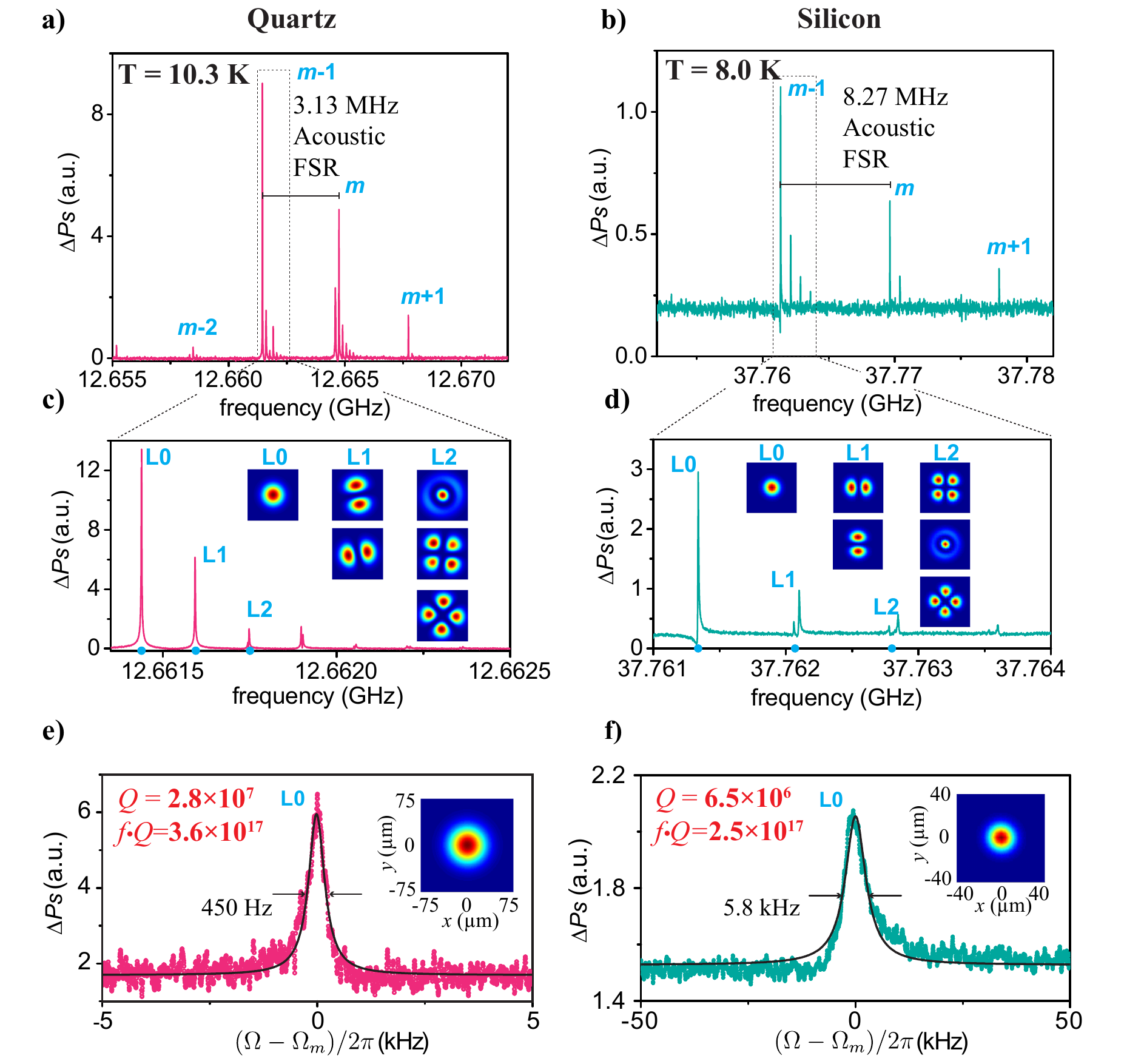}
\caption{\label{fig3} Laser based spectroscopy of standing-wave longitudinal acoustic modes in resonators in z-cut quartz (x-cut silicon) at 10.3 K (8.0 K). Back-scattered probe power ($\Delta P_s$) was recorded as a function of frequency detuning between the pump and the probe light. a)-b) Resonances corresponding to several longitudinal phonon modes with mode number ($m$) separated by the acoustic FSR of 3.13 MHz (8.27 MHz) is observed within the Brillouin-phase matching bandwidth. c)-d) As we zoom in on one acoustic FSR, we find several equally spaced resonances corresponding to higher-order transverse phonon modes (L0, L1, L2, and so on) of the plano-convex geometry. These modes have frequency spacings of 154 kHz (732 kHz) for quartz (silicon) resonators. The simulated frequency spacing (blue dots) accounting for the crystal anisotropy matches very well with the experimental results for resonators in both quartz and silicon. Simulated phonon mode intensity at the plane surface is plotted showing Hermite-Gaussian-like transverse mode profiles. e)-f) Measurement of the fundamental longitudinal mode (L0) at low optical powers reveals a narrow full-width at half-maximum (FWHM) of 450 Hz (5.8 kHz) for a phonon mode at 12.66 GHz (37.76 GHz) in quartz (silicon).}
\end{figure}

\section{Experimental Studies}
%These fabrication results show promising features that hint at phonon cavities that should support long-lived phonons with minimal scattering losses. We probe these phononic resonators using optomechanical transduction at cryogenic temperatures. Next, we briefly explain this optomechanical coupling (further details can be found in Ref.\cite{renninger2017bulk})  
We probe these phononic resonators using laser-based optomechanical spectroscopy at cryogenic temperatures. We engineer the phonon modes of our resonators to be Brillouin-active, such that optical forces generated by photoelastic response of the material permits light-sound coupling. Through a phase-matched Brillouin-like interaction, laser fields can couple to high frequency longitudinal acoustic phonon modes over a finite bandwidth near Brillouin frequency, $\Omega_s =2 \omega_p v_a/v_o$. Here, $v_a$($v_o$) is the speed of sound (light) in the bulk medium, and $\omega_p$ is the frequency of the pump light. The coupling bandwidth due to phase-matching constraints is approximately twice the acoustic free spectral range (FSR) of $v_a/2t$ (typically 1-50 MHz). Through this interaction, stimulated energy transfer between the counter-propagating pump and probe-waves coincides with generation of coherent phonons within the crystal (See Fig. \ref{fig1} c). Therefore, by sweeping the frequency detuning between these laser-fields, we can perform high frequency phonon spectroscopy; multiple resonances corresponding to the standing wave longitudinal phonon modes centered around $\Omega_s$ appear in the energy transfer spectrum and the linewidth of these resonances is determined by the phonon dissipation rate $\Gamma/2\pi$. This technique is very versatile as it permits us to perform phonon spectroscopy in practically any transparent crystalline medium (with or without piezoelectric response). Moreover, since this technique is non-invasive, it permits us to perform rapid spectroscopy on arrays of chip-scale phonon cavities.

We use counter-propagating pump and probe light derived from the same laser source (Pure-Photonics PCL200) at 1549 nm to obtain energy transfer spectra at cryogenic temperatures. The frequency of the pump light is fixed at the laser frequency while the detuning between the pump and the probe light ($\Omega = \omega_p-\omega_s$) is swept through the Brillouin frequency ($\Omega_s$). The magnitude of energy transferred from pump light to probe light ($\Delta P_s$) is recorded as a function of $\Omega$ (details on optomechanical coupling and measurement setup can be found in Ref. \cite{renninger2017bulk}).

Through this nonlinear spectroscopy, we identify families of narrow resonances corresponding to longitudinal standing wave acoustic modes near the Brillouin frequency of 12.66 GHz (37.76 GHz) for resonators in quartz (silicon) at 10.3 K (8.0 K) (See Fig.\ref{fig3} a-b). These modes families ($m$, $m+$1 and so on) are separated by the acoustic FSR of 3.13 MHz (8.27 MHz) as expected for 1 mm thick z-cut quartz (500 $\upmu$m thick x-cut silicon). As we zoom in on one acoustic FSR, we see multiple equally spaced resonances separated by 154 kHz (761 kHz) for resonators in quartz (silicon). These resonances (L1, L2, and so on) are higher order transverse phonon modes (Hermitte-Gaussian-like modes) of the plano-convex geometry. The frequency spacing for the higher order modes agrees well with the analytically calculated spacing of 155 kHz (731 kHz) (See Supplementary Note for details).

We also compared our experimental results (and the analytical theory) with the numerical beam propagation method (of Ref. \cite{renninger2017bulk}), which accounts for the full anisotropy in the elastic tensor. This numerical method was used to calculate the mode spacing (blue dots in Fig.\ref{fig3} c-d) as well as the acoustic mode profiles seen in Fig.\ref{fig3} c-f.
Note that we compare the relative mode spacing, as uncertainties in the elastic constants at cryogenic temperatures prevent us from determining the absolute mode indices (i.e., the precise number of overtones). Numerical calculations revealed good agreement with the experimental mode spacing for resonators fabricated from both quartz and silicon (See Fig.\ref{fig3} c-d). The finer resonance structures, such as the observed splitting ($\sim$40 kHz) of higher order modes (L1, L2) in silicon (See Fig.\ref{fig3} d), are consistent with small asymmetries in the convex surface of the resonator; this was identified using a combination of surface profile measurements and simulations.

In general, the most tightly confined fundamental mode (L0) within a given mode family ($m$) is expected to exhibit the lowest diffraction (and anchoring) losses. To explore the performance of this on-chip phononic resonator, we now focus on the fundamental phonon mode (L0). To ensure that we converge on the intrinsic phonon linewidth (i.e., eliminating possible line-shape distortions arising from acoustic non-linearities) we greatly reduce the laser powers as we perform the high resolution measurements seen in Fig \ref{fig3} e-f. At such powers, we estimate phonon intensities of around 10 Wm$^{-2}$ (220 Wm$^{-2}$) in quartz (silicon). Analysis of the line-shapes in Fig \ref{fig3} e-f revealed linewidth (full-width at half-maximum) of 450 Hz (5.8 kHz) corresponding to an ultra-high-$Q$ of $2.8 \times 10^7$ ($6.6\times10^6$) for 12.7 GHz (37.8 GHz) phonon mode in resonators on-chip  in z-cut quartz (x-cut silicon). This large $f \! \cdot \! Q$- product of $3.6\times10^{17}$ ($2.5\times10^{17}$) obtained for the on-chip resonator in quartz (silicon) is comparable to $f \! \cdot \! Q$- products obtained in macroscopic bulk-crystalline resonators \cite{galliou2013extremely,renninger2017bulk}. 

\section{Discussion}
These techniques to fabricate and characterize on-chip phononic resonators could be extended to perform scalable studies of phonon dissipation, surface interactions, and defects/impurities in a broad class of materials. As for technological applications, on-chip plano-convex phononic resonators of the same form factor discussed here can be coupled to superconducting qubits \cite{Chu199}. While we have designed our phononic devices to have high $Q$ at $>$10 GHz corresponding to the Brillouin frequency, these design strategies can be used to create chip-scale cHBAR that supports low-loss phonon modes in frequency ranges (5-10 GHz) relevant to these superconducting circuits. Since the lifetimes of these phonon modes can be much longer than the superconducting qubit lifetimes, such resonators could enable storage of quantum information on-chip \cite{Chu199}. 

In conclusion, these results lay a foundation for versatile non-invasive materials spectroscopy techniques and new device strategies to benefit quantum information. We have developed simple microfabrication techniques to fabricate arrays of phononic resonators on-chip and precisely control their geometry. Through laser-based spectroscopy of 12.7 GHz phonon modes of cHBAR in quartz at cryogenic temperatures, we have demonstrated $f \! \cdot \! Q$-products ($3.6 \times 10^{17}$) that are comparable to the world class $f \! \cdot \! Q$-products obtained on centimeter-scale BAW quartz resonators \cite{renninger2017bulk}. Furthermore, our demonstration of a comparable $f \! \cdot \! Q$-product ($2.5 \times 10^{17}$) in silicon at record high frequency of 37.8 GHz suggests silicon as a great platform to support long-lived high frequency mechanical excitations. Lastly, there is a path to efficiently accessing these high-$Q$ phonons with both light and microwave using optomechanical and electromechanical techniques. Therefore, these chip-scale cHBAR systems could potentially enable coherent information transfer from microwave to optical domain. Finally, it is intriguing to consider the possibility of utilizing these resonators in silicon to realize novel solid-state quantum devices consisting of spin qubits coherently coupled to both photons and phonons \cite{soykal2011sound,ruskov2012coherent}.

\section{Acknowledgements}
We acknowledge funding support from ONR YIP (N00014-17-1-2514), NSF MRSEC (DMR-1119826) and the Packard Fellowship for Science and Engineering. This research was also supported by the U.S. Army Research Office (W911NF-14-1-0011). Facilities use was supported by the Yale SEAS cleanroom, the Yale West Campus cleanroom and the Yale Institute for Nanoscience and Quantum Engineering (YINQE). The authors thank Luigi Frunzio, Eric Kittlaus, Nils Otterstrom and Shai Gertler for helpful discussions and feedback. We greatly appreciate the support of our cleanroom staff: Chris Tillinghast, James Agresta and Min Li. The authors of this paper are contributors to patent application no. 62/465101 related to Bulk Crystalline Optomechanics and patent application no. 62/465101 related to Techniques for Coupling Qubits to Acoustic Resonators and Related Systems and Methods, which were submitted by Yale University.

\section{Supplemenatry Material}
Refer to this section for the complete description of stability criteria for cHBAR, anisotropy parameter, and anchoring loss estimates.

    \subsection{Anisotropy parameter and Stability Criterion}
    In this section, we derive the stability criterion for longitudinal acoustic modes in anisotropic crytalline medium along certain crystalline axes. Let us first look at longitudinal modes propagating along $x$-cut silicon. We start with the Christoffels' equation for elastic wave propagation in an anisotropic medium \cite{royer2000elastic}
    \begin{equation} \label{christoeffelsequation}
        \rho \frac{\partial^2 u_i}{\partial t^2} = c_{ijlm} \frac{\partial^2 u_m}{\partial x_j \partial x_l},
    \end{equation}
    where $u_i$ is the $i^{\text{th}}$ component of the acoustic displacement vector, $c_{iklm}$ is the elastic constant tensor and $\rho$ is the density of the medium. Using the symmetry property of the elastic tensor in cubic crystal Eq. (\ref{christoeffelsequation}) reduces to 
    \begin{align}
      \rho \ddot{u}_1 &= c_{11} u_{1,11}+ c_{44} u_{1,22}+c_{44} u_{1,33}+(c_{12}+c_{44})(u_{2,12}+u_{3,13}), \\
      \rho \ddot{u}_2 &= c_{44} u_{2,11}+ c_{11} u_{2,22}+ c_{44} u_{2,33} + (c_{44}+c_{12}) (u_{1,12}+u_{3,23}), \\
      \rho \ddot{u}_3 &= c_{44} u_{3,11} + c_{44} u_{3,22} + c_{11} u_{3,33} + (c_{12}+c_{44}) (u_{1,13}+u_{2,23}),
    \end{align}
    where $c_{\alpha \beta}$ is the reduced elastic tensor coefficient, and $u_{m,jl} \equiv (\partial^2 u_m)/(\partial x_j \partial x_l)$. We see that the diffraction terms ($c_{44} u_{1,22}, c_{44} u_{1,33}$) for the longitudinal acoustic wave propagating along $\hat{x}$ is symmetric about $x$-axis. So for simplicity, we consider 2D acoustic beam propagation along $x-y$ plane (i.e. ignore terms related to the displacement $u_3$). This results in the following coupled differential equation for the acoustic displacement fields
    \begin{align} \label{coupled2D1}
         \ddot{u}_x &= v_l^2 u_{x,xx}+v_t^2 u_{x,yy}+ \gamma_1^2 u_{y,xy},\\\label{coupled2D2}
         \ddot{u}_y &= v_t^2 u_{y,xx}+v_l^2 u_{y,yy}+ \gamma_1^2 u_{x,xy},
    \end{align}
    where $v_l = \sqrt{c_{11}/\rho}, v_t = \sqrt{c_{44}/\rho},$ and $\gamma_1 = \sqrt{(c_{12}+c_{44})/\rho}$.
    We now consider a longitudinal wave propagating along $x$-direction i.e. $\textbf{u} (\textbf{r},t) = \textbf{A}(\textbf{r})e^{-i(k_ox-\Omega t)}$, where $\textbf{A}(\textbf{r}) = A_x (x,y) \hat{x}+ A_y(x,y) \hat{y},$ and $k_o = \Omega/v_l.$ For paraxial beam propagation along $x$, we make the slowly varying envelope approximation (i.e. $\partial^2 A_i/\partial x^2 \ll ko^2 A_i , i = x,y$) to obtain the following coupled equations from Eqs. (\ref{coupled2D1}-\ref{coupled2D2})
    \begin{align}
        -k_o^2v_l^2 A_x &= -k_o^2v_l^2 A_x -2ik_ov_l^2 \frac{\partial A_x}{\partial x}+v_t^2 \frac{\partial^2 A_x}{\partial y^2} - i \gamma_1^2 k_o \frac{\partial A_y}{\partial y}, \\
        -k_o^2v_l^2 A_y &= -k_o^2 v_t^2 A_y - 2ik_o v_t^2 \frac{\partial A_y}{\partial x}+v_l^2 \frac{\partial^2 A_y}{\partial y^2} - i \gamma_1^2 k_o \frac{\partial A_x}{\partial y}.
    \end{align}
    To solve these equations, we Fourier transform \textbf{A}(\textbf{r}) to $k$-space. So using $A_x(x,y) = 1/\sqrt{2\pi} \int dk_y \tilde{A}_x (x,k_y) e^{ik_y y}$ and $A_y(x,y) = 1/\sqrt{2\pi} \int dk_y \tilde{A}_y (x,k_y) e^{ik_y y}$, we get following first order differential equations 
    \begin{align} \label{Aeq1}
        \frac{\partial \tilde{A}_x}{\partial x} + i p \tilde{A}_x + i q \tilde{A_y}=0,\\ \label{Aeq2}
        \frac{\partial \tilde{A}_y}{\partial x} + i r \tilde{A}_y + i s \tilde{A_x}=0,
    \end{align}
    where $p = -(v_t^2 k_y^2)/(2v_l^2 k_o), q = \gamma_1^2k_y/(2v_l^2), r = -(k_o^2(v_t^2-v_l^2)+v_l^2k_y^2)/(2v_t^2k_o),$ and $s = \gamma_1^2k_y/(2v_t^2).$ Applying $(\partial_x +ir)$ on Eq. (\ref{Aeq1}) and making the paraxial approximation, we finally get
    \begin{align} \label{finalA}
        \frac{\partial \tilde{A}_x(x,k_y)}{\partial x} + i \frac{rp-qs}{p+r} {\tilde{A}_x (x,k_y)}=0.
    \end{align}
    Assuming $k_y\gg k_o$ in the paraxial limit we get
    \begin{align}
        \frac{rp-qs}{p+r} \simeq \frac{k_o(v_l^2v_t^2 -v_t^4+\gamma_1^4)}{2v_l^2 (v_l^2-v_t^2)} \left(\frac{k_y}{k_o}\right)^2.
    \end{align}
    We can now re-write Eq. (\ref{finalA}) as
    \begin{align} \label{paraxialEq}
    2k^{\prime}i\frac{\partial \tilde{A}_x}{\partial x} = -k_y^2 \tilde{A}_x,
    \end{align}
    where $k^{\prime} = k_o \chi$, and the ``anisotropy-parameter", $\chi$ is given by
    \begin{align}
        \chi = \frac{v_l^2(v_l^2-v_t^2)}{v_l^2v_t^2-v_t^4+\gamma_1^4}.
    \end{align}
    Note that Eqn. (\ref{paraxialEq}) for acoustic wave propagation in the paraxial limit, is similar to the paraxial approximation to the scalar wave equation for electromagnetic field \cite{saleh1991fundamentals}. However, for the acoustic wave propagation in the paraxial limit there is an additional factor of $\chi$ in the propagation constant (i.e. $k^{\prime} = k_o \chi$). While we have determined $\chi$ analytically, it is also possible to obtain $\chi$ numerically by fitting a quadratic function to the slowness surfaces \cite{newberry1989paraxial}. It is well known from optics that Gaussian beams satisfy the paraxial approximation to the wave equation \cite{siegman1986lasers}. Therefore, assuming acoustic field polarized along $x$ (i.e. $\textbf{u} (\textbf{r},t) = A(x,y)e^{-i(k_ox-\Omega t)} \hat{x},$ with initial acoustic field at $x=0$ as $A(x=0,y)$ = $A_o \text{exp}(-y^2/w_o^2)$, we can solve Eq. (\ref{paraxialEq}) to get
    \begin{align}
        \textbf{u}(\textbf{r},t) &= A_o \hat{x} \frac{w_o}{w^{\prime}(x)} \text{exp}\left( -\frac{y^2}{w^{\prime}(x)^2}\right) \text{exp}\left(-ik_o\frac{y^2}{2R^{\prime}(x)} + i\psi^{\prime}(x)\right) \text{exp}\left(-i(k_ox-\Omega t \right)),
    \end{align}
    where
    \begin{align*}
        &k^{\prime} = k_o\chi = \frac{2\pi}{\lambda_\text{ph}}\chi \text{ for phonon with wavelength }
        \lambda_{\text{ph}},\\
        &w_o \text{ is the acoustic waist radius at } x=0, \\ 
        &w^{\prime}(x) = w_o \sqrt{1+\left(\frac{x}{x_\text{R}^{\prime}}\right)^2} \text{ is the acoustic waist radius at } x, \\
        &x_{\text{R}}^{\prime} = \frac{\pi w_o^2}{\lambda_{ph}}\chi  \text{ is the acoustic Rayleigh length}, \\
        &R^{\prime}(x) = \frac{1}{\chi}\left(x+\frac{{x^{\prime}_{\text{R}}}^2}{x}\right) \text{ is the radius of curvature of the acoustic beam's wavefronts at x}, \\
        & \psi^{\prime}(x)= \text{arctan}\left(\frac{x}{x^{\prime}_{\text{R}}}\right) \text{ is the acoustic Gouy phase at } x. 
    \end{align*}
    Now that we have determined propagation equations for the Gaussian acoustic wave, we can perform stable Fabry-P{\'e}rot resonator analysis for the acoustic cavity similar to that for a two-mirror optical cavity \cite{siegman1986lasers}. 
    \begin{figure*}
    \includegraphics[width=0.5\textwidth]{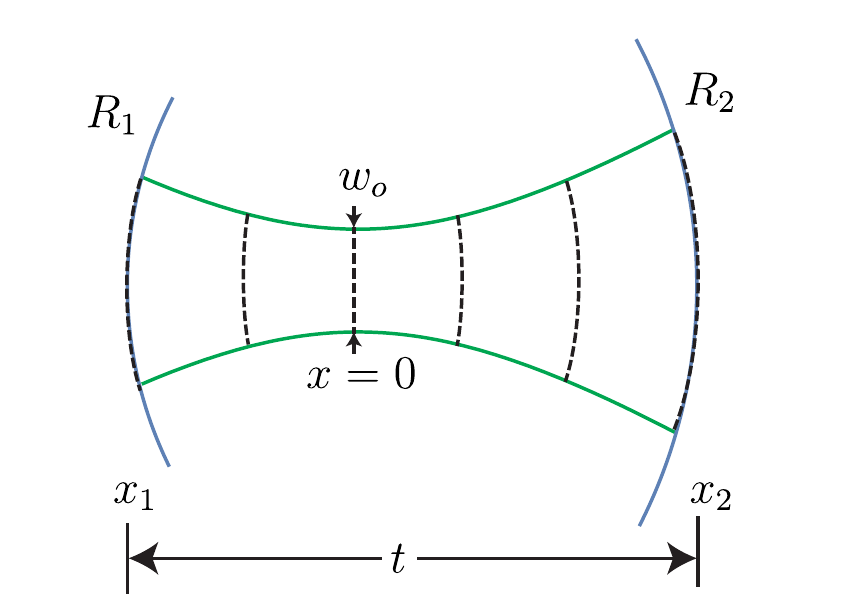}
    \caption{\label{fig4} Phonon cavity of length $t$ with two surfaces with radii of curvature $R_1$ and $R_2$. }
    \end{figure*}
    Given two surfaces with radius of curvatures $R_1$ and $R_2$ with spacing $t$ between, we need to find a Gaussian beam that periodically refocuses upon each round trip. Let us assume that this Gaussian beam with an initially unknown spot size $w_o$ is at an initially unknown location such that the reflecting surfaces are at distances $x_1$ and $x_2$ away (See Fig. \ref{fig4}). For periodic refocusing, the acoustic beams' radius of curvatures $R^{\prime}(x)$ need to match with radius of curvatures of the two mirrors. This gives us three equations
    \begin{align} \label{eqR1}
        R^{\prime}(x_1) &=\frac{1}{\chi}\left( x_1+\frac{{x_\text{R}^{\prime}}^2}{x_1}\right) = R_1, \\ \label{eqR2}
      R^{\prime}(x_2) &=\frac{1}{\chi}\left( x_2+\frac{{x_\text{R}^{\prime}}^2}{x_2}\right) = R_2, \\ \label{eqt}
      t &= x_2-x_1.
    \end{align}
    
    We now invert Eqns. (\ref{eqR1}-\ref{eqt}) to find the Gaussian beam parameters $x^{\prime}_\text{R}$, $x_1$ and $x_2$ in terms of $R_1$, $R_2$ and $t$. So, if we define acoustic resonator ``$g$-parameter" as follows
    \begin{align}
        g_1 = 1- \frac{t}{\chi R_1} \text{ and }  g_2 = 1-\frac{t}{\chi R_2}  ,
    \end{align}
    we can find the Gaussian beam parameters in terms of $g_1$, $g_2$ and $t$. The Rayleigh length for this trapped Gaussian beam is given by
    \begin{align} \label{eqrayleighac}
        {x_\text{R}^{\prime}}^2 = \frac{g_1g_2(1-g_1g_2)}{(g_1+g_2-2g_1g_2)^2}t^2.
    \end{align}
    The locations of the curved surfaces with respect to the Gaussian beam waist are
    \begin{align}
        x_1 &= \frac{g_2(1-g_1)}{g_1+g_2-2g_1g_2}t,\\
        x_2 &= \frac{g_1(1-g_2)}{g_1+g_2-2g_2g_2}t.
    \end{align}
    From Eq. \ref{eqrayleighac} and the definition of the Rayleigh length derived earlier, we can obtain the Gaussian beam waist size at $x=0$ and $x=t$  
    \begin{align} \label{eqbeamwaist}
        w_o^2 &= \frac{t \lambda_{\text{ph}}}{\pi \chi} \sqrt{\frac{g_1g_2(1-g_1g_2)}{(g_1+g_2-2g_1g_2)^2}} ,\\ \label{eqbeamwaist2}
        w_2^2 &= \frac{t \lambda_{\text{ph}}}{\pi \chi} \sqrt{\frac{g1}{g_2(1-g_1g_2)}}.
    \end{align}
    From equations (-\ref{eqrayleighac}) \& (\ref{eqbeamwaist} ) we see that for real and finite solutions to the Gaussian beam parameters and the beam waist size, we see that $0\leq g_1g_2 \leq 1.$ For a plano-convex phononic resonator discussed in this paper, $R_1= \infty$ so, $g_1= 1$ and $x_1=0$.
    
    We can also analytically calculate the frequency spacing between the higher-order transverse modes, because in the paraxial limit higher order Hermite-Gaussian modes are also solutions to the paraxial equation in Eq. (\ref{paraxialEq}).
    \begin{align}
        u_n(x,y)& = A_o \hat{x} H_n\left( \frac{\sqrt{2}y}{w^{\prime}(x)}\right) \times \\
        & \text{exp}\left( -\frac{y^2}{w^{\prime}(x)^2}\right) \text{exp}\left(-ik_o\frac{y^2}{2R^{\prime}(x)} + i(n+1/2)\psi^{\prime}(x)\right) \text{exp}\left(-ik_ox\right),
    \end{align}
    where $H_n$s are the Hermite polynomials of order $n$ and are also solutions to Eq. (\ref{paraxialEq}). 
    To calculate the higher order mode frequency spacing we look at the total round trip phase shift along the cavity axis (i.e. $y=0$), which must be an integer multiple of $2\pi$. This gives us the following equation 
    \begin{align}
    -k_ot+(n+1/2)(\psi^{\prime}(t)-\psi^{\prime}(0)) &= \pi \\
    \frac{-\Omega_n}{v_l} t + (n+1/2)\psi^{\prime}(t) &= \pi
    \end{align}
    So the higher-order transverse mode spacing is given by
    \begin{align} \label{eqhigherordermodespacing}
        \Delta f= \frac{\Omega_n-\Omega_{n-1}}{2\pi} = \frac{1}{2\pi}\text{arctan}\left(\frac{t}{x^{\prime}_{\text{R}}}\right) \frac{v_l}{t}, 
    \end{align}
    with the Rayleigh length given by Eq. (\ref{eqrayleighac}).
    
    We can now calculate the acoustic beam waist size $w_o$ and the higher-order mode frequency spacing analytically for plano-convex phonon cavities in 0.5 mm thick $x$-cut silicon, with $R_2= 13.3$ mm. Using $\rho = 2329 $ kg m$^{-3}$, $c_{11} = 165.6$ GPa , $c_{44} = 79.5 $ GPa, $c_{12} = 63.9$ GPa, we find that $v_l = \sqrt{c_{11}/\rho} $ = 8432 ms$^{-1}$, $v_t = \sqrt{c_{44}/\rho}$ = 5843 m s$^{-1}$, $\gamma_1 = \sqrt{(c_{12}+c_{44})/\rho}$ = 7847 m s$^{-1}$. We find that $\chi = 0.5202$ and for phonons with frequency 37.76 GHz, using Eq. (\ref{eqbeamwaist}) we get $w_o= 15.6 \  \upmu $m at $x=0$. The Rayleigh length using Eq. (\ref{eqrayleighac}) is $x^{\prime}_{\text{R}} = 1.791$ mm and the higher-order transverse mode frequency spacing using Eq. (\ref{eqhigherordermodespacing}) is 731 kHz. Given uncertainty in the elastic constants at low temperature, this predicted frequency spacing agrees well with the experimentally measured frequency spacing of 761 kHz.
    
    For acoustic beam propagation along $z$-cut quartz crystal, a similar approach to that outlined above for silicon can be used to obtain the ``anisotropy parameter'' 
    \begin{align}
        \chi = \frac{v_l^2(v_l^2-v_t^2)}{v_l^2v_t^2-v_t^4+\gamma_1^4},
    \end{align}
    where $v_l= \sqrt{c_{33}/\rho}, v_t = \sqrt{c_{44}/\rho}$ and $\gamma_1 = \sqrt{(c_{13}+c_{44})/\rho}$. 
    For plano-convex phonon cavities in 1 mm thick $z$-cut quartz with $R_2 = 65$ mm, we can calculate the acoustic beam size $w_o$ and the higher-order mode spacing analytically.  Using $\rho = 2648 $ kgm$^{-3}$, $c_{33} = 105.75$ GPa , $c_{44} = 58.2 $ GPa, $c_{13} = 11.91$ GPa, we find that $v_l = \sqrt{c_{33}/\rho} $ = 6319 m s$^{-1}$, $v_t = \sqrt{c_{44}/\rho}$ = 4688 m s$^{-1}$, $\gamma_1 = \sqrt{(c_{13}+c_{44})/\rho}$ = 5146 m s$^{-1}$. We find that $\chi = 0.6545$ and for phonons with frequency 12.66 GHz, using Eq. (\ref{eqbeamwaist}) we get $w_o= 39.5 \ \upmu $m at $x=0$. The Rayleigh length using Eq. (\ref{eqrayleighac}) is $x^{\prime}_{\text{R}} = 6.44$ mm and the higher-order transverse mode frequency spacing using Eq. (\ref{eqhigherordermodespacing}) is 155 kHz. This predicted frequency spacing agrees well with the experimentally measured frequency spacing of 154 kHz.
    
    \subsection{Anchoring loss estimates}
    In this section, we calculate the $Q$-factor limit for the fundamental longitudinal acoustic modes (L0) if we assume all the energy outside the diameter ($d$) of the convex surface is lost due to absorption. We will call this our anchoring/clamping loss.
    
    First, we can use Eqns. (\ref{eqbeamwaist}- \ref{eqbeamwaist2}) to see that the fundamental longitudinal mode waist do not vary significantly along the $x$-axis since the Rayleigh range for the cavities in quartz ($R_2$ = 65 mm, $t=$ 1 mm) and silicon ($R_2=$ 13.3 mm, $t =$ 0.5 mm) is much larger than the crystal thickness $t$. In this case, the fundamental Gaussian mode has a displacement profile 
    \begin{align} \label{phononfield}
        u(x,y,z) = u_o \text{exp}\left(-\frac{(y^2+z^2)}{w_o^2}\right) \text{cos}\left(\frac{m\pi x}{t}\right)
    \end{align}
    
    The acoustic energy, $E \propto \int_V dV \ |u(\textbf{r})|^2$, the fraction of energy that lives outside the convex surface of diameter $d$, using Eq. (\ref{phononfield}) is simply given by
    \begin{align}
        \frac{E_\text{out}}{E_\text{tot}}= e^{\frac{-d^2}{2w_o^2}}.
    \end{align}
    
    If all this energy was lost due to absorption per-round trip, the lifetime, $\tau$, of the acoustic mode in presence of this absorption (ignoring all other intrinsic loss mechanisms) is given by
    \begin{align}
        \tau = -\frac{2 t}{v_l}\frac{1}{\text{ln}(R)},
    \end{align}
    where $R = 1-e^{\frac{-d^2}{2w_o^2}}$, and $v_l$ is the longitudinal acoustic velocity. So the Q-factor limit due to this type of loss would be
    \begin{align}
        Q_\text{achoring} &= \frac{f}{\Delta f} \\
        &= {2\pi \tau f} \\
        &= -\frac{4 \pi t f}{v_l} \frac{1}{\text{ln}\left(1-e^{\frac{-d^2}{2w_o^2}}\right)}.
    \end{align}
    So for $d/w_o$ = 5 for 12.7 GHz mode in 1 mm thick quartz, we estimate a $Q$-factor limit of 6.8 billion due to this type of clamping loss. For $d/w_o$ = 5 for 37.8 GHz mode in 0.5 mm thick silicon, we estimate a $Q$-factor limit of 15 billion. 
    
\bibliographystyle{apsrev4-1} %apsrev4-1

\bibliography{mybib}

\end{document}